\documentclass[prl,twocolumn,superscriptaddress,nofootinbib]{revtex4-2}
\usepackage{amsmath}
\usepackage{amsfonts}
\usepackage{amssymb}
\usepackage{mathtools}
\usepackage{graphicx}
\usepackage[colorlinks]{hyperref}
\usepackage[capitalize]{cleveref}
\usepackage{color}
\usepackage{enumitem}

\usepackage{yypreamble}
\newcommand{\pardash}[1]{\paragraph{#1 ---}}

\newcommand{\pc}{p_{\rm c}}
\newcommand{\pcest}{\overline{\pc}}
\newcommand{\setU}{\mathbf{U}}
\newcommand{\setM}{\mathbf{M}}
\newcommand{\setm}{\mathbf{m}}
\newcommand{\F}{F}
\newcommand{\C}{\mathcal C}
\newcommand{\en}{\epsilon^{(n)}_{D}}
\newcommand{\nD}{n_{D}}

\begin{document}

\title{Detecting Measurement-Induced Entanglement Transitions With Unitary Mirror Circuits}

\author{Yariv Yanay}
\email{yanay@umd.edu}
\affiliation{Laboratory for Physical Sciences, 8050 Greenmead Dr., College Park, MD 20740, USA}
\affiliation{Department of Physics, University of Maryland, College Park, MD 20742, USA}
\author{Brian Swingle}
\affiliation{Department of Physics, Brandeis University, Waltham, Massachusetts 02453, USA}
\author{Charles Tahan}
\affiliation{Department of Physics, University of Maryland, College Park, MD 20742, USA}

\begin{abstract}
Monitored random circuits, consisting of alternating layers of entangling two-qubit gates and projective single-qubit measurements applied to some fraction $p$ of the qubits, have been a topic of recent interest. In particular, the resulting steady state exhibits a phase transition from highly correlated states with “volume-law” entanglement at $p<\pc$ to localized states with “area-law” entanglement at $p>\pc$. 
It is hard to access this transition experimentally, as it cannot be seen at the ensemble level. Naively,  to observe it one must repeat the experiment until the set of measurement results repeats itself, with likelihood that is exponentially small in the number of measurements.
To overcome this issue, we present a hybrid quantum-classical algorithm which creates a matrix product state (MPS) based “unitary mirror'' of the projected circuit. Polynomial-sized tensor networks can represent quantum states with area-law entanglement, and so the unitary mirror can well-approximate the experimental state above $\pc$ but fails exponentially below it. The breaking of this mirror can thus pinpoint the critical point.
We outline the algorithm and how such results would be obtained. We present a bound on the maximum entanglement entropy of any given state that is well-represented by an MPS, and from the bound suggest how the volume-law phase can be bounded. We consider whether the entanglement could similarly be bounded from below where the MPS fails. Finally, we present numerical results for small qubit numbers and for monitored circuits with random Clifford gates.
\end{abstract}
\maketitle

Monitored random circuits, consisting of alternating layers of random entangling unitaries and local measurement operations on a randomly chosen subsets $p$ of all qubits, have recently come into focus as a probe into the interplay of unitary and non-unitary evolution, quantum error correcting codes, and more \cite{Skinner2019, Szyniszewski2020, Choi2020, Gullans2020, Gopalakrishnan2021, Ippoliti2021, Bao2024}. These circuits have been shown to exhibit a phase transition between long range, volume-law entanglement between the qubits at $p<\pc$ and a local, area-law entanglement phase at $p>\pc$, for some critical measurement density $\pc$ \cite{Zabalo2020}. Significantly, this is not a property of the entanglement of the statistical ensemble of states; it is only seen if one averages the entanglement of each individual outcome, characterized by the circuit and the record of measurement results \cite{Friedman2023}. This makes it difficult to experimentally access these properties, as generally the number of repetitions required to recreate same state twice grows exponentially with the number of measurements. 

\begin{figure}[thbp] 
   \centering
   \includegraphics[width=\columnwidth]{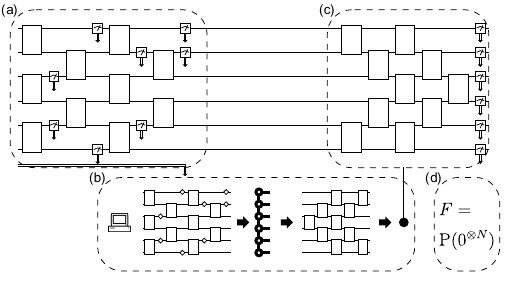} 
   \caption{Outline of our proposed scheme. {\bf(a)} A randomly monitored circuit is run on a quantum device generating a state $\ket{\psi}$.
   	{\bf(b)} The gate parameters and measurement results are fed into a classical processor, generating a deterministic circuit where measurements have been replaced with projection operators (marked as diamonds). We classically calculate an MPS $\ket{\psi_{D}}$ approximating the output of this circuit, and from it a “unitary mirror'' circuit that generates it, $\mathcal C_{D}\ket{0^{\otimes N}} = \ket{\psi_{D}}$.
	{\bf(c)} Finally, we apply the inverted mirror $\mathcal C_{D}\dg$ to $\ket{\psi}$ and measure all qubits.
	{\bf(d)} The probability of finding the all-zero state is proportional to the overlap $\abs{\braket{\psi_{D}}{\psi}}^{2}$.
   }
   \label{fig:sketch}
\end{figure}

A variety of lateral approaches have been suggested to overcome this difficulty. One tack has been to apply a form of steering, or preselection, to direct the system into a single state in the entangled regime \cite{Buchhold2022, Iadecola2023, Piroli2023, Sierant2023}. While this allows for experimental implementation, the dynamics and universality regime generally diverge from those of undirected MIPT formulations. Techniques looking to probe the undirected form usually rely on some classical calculation that attempts to recreate the quantum dynamics of the circuit, and correlate the final measurement with that calculation \cite{Gullans2020a, Lee2022, Noel2022, Hoke2023, Dehghani2023, Garratt2023, Li2023a}. 
In principle, for any instance of a monitored circuit, consisting of an initial state, a set of unitary operations and measurement points, and a record of measurement results, the final state of the system is deterministic, and could be obtained by a classical computer.
In practice, this generically requires calculating the unitary matrix of a many-qubit operation, and thus one has simply shifted the exponential requirement from the experimental regime into the computational. 
To avoid this overhead, experimental approaches generally use an approximation of the state to extract the entropy, introducing errors or uncertainty to the result.

Here, we propose an alternate approach for the use of classical calculations, leveraging the known limitations of these approximations as a diagnostic tool in itself by making use of  tensor network representations of quantum states \cite{Orus2014, Cirac2021}. 
Tensor networks can be used to represent a subset of many-body quantum states with linear, rather than exponential, overhead in the number of qubits. Likewise, the translation of an instance of a monitored circuit into a unitary-only equivalent only requires a number of operations linear in the number of qubits and the depth of the circuit. The downside of tensor networks is that they cannot accurately represent states with volume-law entanglement. We turn this deficiency into a diagnostic tool: as the system crosses from area-law entropy to volume law entropy states, we expect to see a similar transition in the overlap between the real state generated by the monitored circuit and its TN-generated copy. By measuring this fidelity, as outlined in \cref{fig:sketch}, we can experimentally access to the transition point and measure $\pc$.

\pardash{Protocol}
We consider a chain of $N$ qubits and an $L$-layer circuit $\C$. The circuit is composed of a set of gates, $\setU = \{\hat U_{\ell,2q} | 1\le\ell\le L, 1\le q\le \lfloor N/2\rfloor\}$, sampled out of some set of two-qubit unitaries $\hat U_{\ell,q}\in\mathcal U$, and measurement points, $\setM = \{M_{\ell}|1\le\ell\le L\}$, where $M_{\ell}\subset\{1,\dotsc,N\}$, $\abs{M_{\ell}} = pN$ for some measurement rate $p$.
At each odd (even) layer $\ell$, we apply $\hat U_{\ell,q}$ between each even qubit $2q$ and its preceding (following) neighbor, $2q+\p{-1}^{\ell}$. After this, we measure each qubit $q\in M_{\ell}$ recording the results into an $L\times pN$ matrix $\setm$. The circuit generates a final state $\ket{\psi\br{\setU, \setM, \setm}}$.

To probe entropy properties, we extend the circuit as follows. We run $\C$ as given above, generate $\ket{\psi}$, and keep it in quantum memory.
Then $\setU$, $\setM$, and $\setm$ are sent to a classical computer, which calculates an approximate final state $\ket{\psi_{D}}$ using a matrix product state (MPS) decomposition with bond dimension $D$ \cite{Cirac2021}. 
We convert it \cite{Schon2005, Ran2020,supplemental} into a “unitary mirror'', a unitary-only circuit that when applied to an all-zero states, generates $\C_{D}\ket{0^{\otimes N}} = \ket{\psi_{D}}$.
The quantum device then applies the inverse of the mirror to the state, to generate $\ket{\phi_{D}} = \C_{D}\dg\ket{\psi}$.
Finally, we measure the probability of finding all qubits in the zero state, to obtain the overlap
\begin{equation}\begin{split}
\F_{D}\br{\setU, \setM, \setm} 
	& = \abs{\braket{0^{\otimes N}}{\phi_{D}\br{\setU, \setM, \setm}}}^{2} 
	\\ & = \abs{\bra{0^{\otimes N}}\C_{D}\dg\br{\setU, \setM, \setm}\ket{\psi\br{\setU, \setM, \setm}}}^{2}
	\\ & = \abs{\braket{\psi_{D}\br{\setU, \setM, \setm}}{\psi\br{\setU, \setM, \setm}}}^{2}.
\end{split}\end{equation}
By repeating this process, we measure the mirror fidelity averaged over all possible outcomes $\setm$.

\begin{figure}[t] 
   \centering
   \includegraphics[width=\columnwidth]{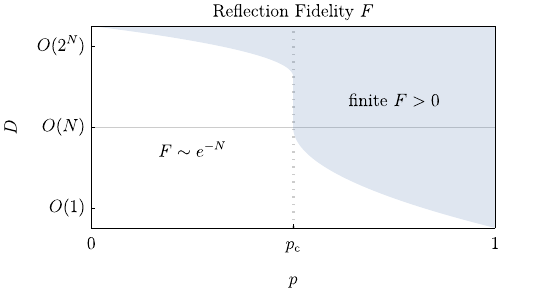} 
   \caption{Expectations from the mirror fidelity of a monitored circuit of $N$ qubits, sketched out as a function of the measurement probability $p$ and the bond dimensions $D$ (in log scale). Where the graph is shaded we expect some finite fidelity, while where it is left white we expect the fidelity to be exponentially small in the number of qubits.  
    On the left, for $p<\pc$, we expect the wavefunction to have volume-law entropy behavior and so the bond dimension required to approximate it is exponentially large in the qubit number $N$. For $p>\pc$, where the wavefunction has area-law entropy, a polynomial bond dimension should be sufficient.
   }
   \label{fig:phase}
\end{figure}

How does the behavior of $\F$ vary with $p$? At $p=0$, an MPS with $D\ll 2^{N}$ is not sufficient to approximate the state and $\F$ approaches zero exponentially with $N$. At $p=1$, where $\ket{\psi}$ remains a simple Fock state after each layer, we expect the MPS to perfectly approximate the circuit and find $\F=1$. In between, we may expect it to follow the behavior of the entropy curve: Where the state has area-law or $\log N$ entropy at most, a polynomial $D\sim N^{\gb}$ may be sufficient to generate a sufficiently good approximate state so that $\F\sim 1$; while for volume-law entropy, the MPS will fail to capture the state and we expect the mirror fidelity to drop off exponentially with $N$.
This is sketched out in \cref{fig:phase}.

\pardash{Upper bound on entropy}
To quantify this relation, we consider the Schmidt decomposition of $\ket{\psi}$ to the subsets of qubits $\{1\dotsc n\}$, $\{n+1\dotsc N\}$,
\begin{equation}
\ket{\psi} = \smashoperator{\sum_{i=1}^{2^{\bar n}}}\sqrt{\mu^{(n)}_{i}}\ket{\psi_{1\dotsc n}}\ket{\psi_{n+1\dotsc N}},
\end{equation}
where $\bar n = \min\br{n,N-n}$, $\sum_{i} \mu^{(n)}_{i} = 1$, and $\mu^{(n)}_{i}\ge\mu^{(n)}_{i+1}$.
The Schmidt error \cite{Verstraete2006} is then defined as 
\begin{equation}
\en = \smashoperator{\sum_{i=D+1}^{2^{\bar n}}}\mu^{(n)}_{i}.
\label{eq:defen}
\end{equation}
By construction, the mirror fidelity is bounded by ${\F \le 1 - \max_{n}\gep^{(n)}_{D}}$ \cite{supplemental}. 

Next, consider the Von Neuman entanglement entropy on the subset, $S^{(n)} = -\sum_{i}\mu^{(n)}_{i}\log\mu^{(n)}_{i}$. For a given Schmidt error, the entropy can be maximized by taking $\mu_{i\le D} = (1-\en)/D$, $\mu_{i>D} = \en/(2^{\bar n}-D)$. Thus, given $\en$, the entanglement entropy of the subset $i=1\dotsc n$ is bounded by
\begin{equation}
S^{(n)} \le \bar S({\gep^{(n)}_{D}}) + \p{1-\gep^{(n)}_{D}}\log D + \gep^{(n)}_{D}\log\p{2^{\bar n}-D},
\label{eq:ubSeps}
\end{equation}
where $\bar S\p{\gep} = - \p{1-\gep}\log\p{1-\gep} - \gep\log \gep$.

\begin{figure}[tbp] 
   \centering
   \includegraphics[scale=1]{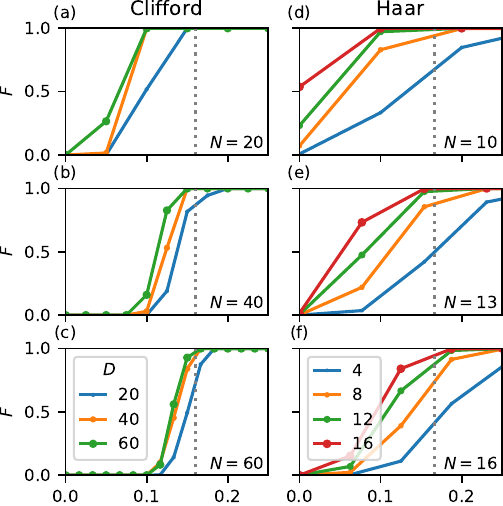} 
   \caption{The average overlap $\F$ of the unitary mirror generated by the MPS for various random circuits. 
   Shown for circuits using {\bf(a-c)} randomly chosen two-qubit Clifford gates and {\bf(d-f)} Haar-random two-qubit gates as entanglers. The dotted gray lines denote the critical point $\pc$ \cite{Sierant2022}.
   Each subplot corresponds to the noted chain length $N$, and each curve corresponds to a different bond dimension $D$. 
   For the Clifford circuits, we  see a clear inflection point at the critical density; for the Haar circuits, we see signs of a similar phenomenon.
   }
   \label{fig:Fidelity}
\end{figure}

\begin{figure}[tbp] 
   \centering
   \includegraphics[scale=1]{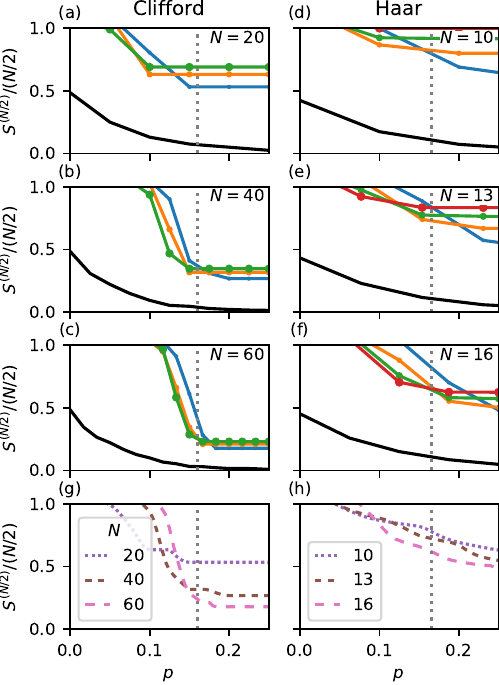} 
   \caption{The upper bound on the value of the volume-law coefficient, derived from overlaps shown in \cref{fig:Fidelity}, as described in the text. 
   Shown for circuits using {\bf(a-c,g)} randomly chosen two-qubit Clifford gates and {\bf(d-f,h)} Haar-random two-qubit gates as entanglers. The dotted gray lines denote the critical point $\pc$ \cite{Sierant2022}.
   {\bf (a-f)} Each subplot corresponds to the noted chain length $N$, and each curve corresponds to a different bond dimension $D$, corresponding to the same plot in \cref{fig:Fidelity}.
   We observe the $\log D$ behavior implied by \cref{eq:ubSF}.
   {\bf(g-h)} We extract an overall curve for each $N$ by taking the minimum value of $S$ over $D$.
   We can see how the bound on the volume law decreases as we increase $N$; this takes the form of the $1/N$ dependence shown in \cref{eq:ubSN}.
   }
   \label{fig:Sub}
\end{figure}

Combining the two bounds \cite{supplemental}, we arrive at our main result. We find that given a state approximated by an MPS of bond dimension $D$ with mirror fidelity $\F$, the entanglement entropy for half the chain is bounded by
\begin{equation}
S^{(N/2)} \le \br{1 + N\p{1-\F}/2}\log 2 + \F\log D.
\label{eq:ubSF}
\end{equation}

This bound allows us to experimentally probe the limits of volume law behavior. Consider the scaling behavior of $\F$ as we increase the number of qubits $N$, and in particular, the scaling of the bond dimension required to keep the fidelity close enough to unity so that $1-\F\le 2/N$. As long as we are able to do so for some $D\le  \p{A N}^{\gb}$, then \cref{eq:ubSF} can be used to rule out any arbitrarily small volume law coefficient by increasing the number of qubits,
\begin{equation}
S^{(N/2)}/N \le \p{2\log 2 + \gb \F\log \br{AN}}/N.
\label{eq:ubSN}
\end{equation}

Importantly, the relation in \cref{eq:ubSF} is linear, and so any averaging over $\F$ corresponds directly to the equivalent averaging for $S$. This is significant as the experimental result we obtain averages over the different circuit outcomes $\setm$. By averaging over $\setU,\setM$, we can find an upper bound on the half-chain entropy and so bound the maximum volume law coefficient. This in turn can be used to find a maximum bound on $\pc$.

We numerically explore this bound in \cref{fig:Fidelity,fig:Sub} for two sets of gates $\mathcal U$ \cite{Johansson2013,Roberts2019,Gidney2021}. For two-qubit gates sampled from the set of Haar-random gates, we consider small chains, up to $N=16$. For $\mathcal U$ consisting of the set of Clifford gates, we consider up to $N=60$. In both cases we consider various bond dimensions up to $D=N$. We observe, in both cases, behavior consistent with the real bound, shown as a black line.

For the Haar-random gates, numerics provide limited evidence, as for the largest qubit number one has $N/2^{N/2} = 16/256=0.0625$, which may be in line with the volume law coefficient near $\pc$. For the Clifford gates, we observe that for $N=60,D=60$ the $\F$ curve appears to be in line with the $D=N$ cut of the sketch in \cref{fig:phase}. Thus, one may find that this is a tight bound on the entropy, allowing us to estimate $\pc$ from the behavior of the curve for large $N$.

To probe $\pc$, we adopt an $\gve$ test, ruling out volume-law behavior whenever the entropy per-unit is below some threshold, $S^{(N/2)}\le \gve N/2$. This can be used to determine an upper bound on on the critical point, $\pcest$, defined as the smallest $p$ where the right-hand side of \cref{eq:ubSF} is smaller than $\gve N/2$.
Rigorously, we then know that for any $p>\pcest$, the volume-law coefficient for the entropy of the state would be smaller than the threshold. 
As noted above, we have numerical evidence that for large $N$, one may find that this upper bound approaches the critical point, $\pcest\to\pc$.

\pardash{Lower bound estimates}
Next we consider whether a small mirror fidelity can be interpreted as a lower bound on the amount of entropy in the system. We can follow a similar procedure to that of the previous section. The upper bound for the entropy is \cite{supplemental}
\begin{equation}
S^{(n)} \ge \bar S({\gep^{(n)}_{D}}) + \gep^{(n)}_{D}\p{1 + \log \br{D\log D}}.
\label{eq:lbSeps}
\end{equation}

Unfortunately, we have found no no easy way to bound $\en$ from below with the fidelity $\F$. Ref.~\onlinecite{Verstraete2006} has shown that there exists a specific MPS construction that has $1 - \sum_{n=1}^{N-1}\gep^{(n)}_{D} \le \sqrt{\F}$, but the time to calculate it from the given circuit scales exponentially with $N$. In addition, this is a very loose bound, as evidenced by the left hand side quickly becoming negative for large values of $N$.

Instead, let us consider the evolution of the approximate wavefunction along the circuit. We obtain $\ket{\psi_{D}}$ by a simple iterative process \cite{supplemental}: we begin with the initial state used in the circuit, $\ket{\psi_{D}^{0}}=\ket{0}^{\otimes N}$. Then, given an approximate state $\ket{\psi_{D}^{\ell-1}}$, we generate $\ket{\psi_{D}^{\ell}}$ by applying the gates and projection operators of layer $\ell$, and then truncating the state back to bond dimension $D$. We can obtain the truncation error of these individual steps as we perform the calculation, which will be smaller than the total Schmidt error. Thus, we can substitute it into \cref{eq:lbSeps} to obtain a lower bound on the entropy of the real state.

The form of \cref{eq:lbSeps} dictates the form of the lower bound we obtain this way. As $\gep^{(n)}_{D}\le 1$, to get better bounds we need to increase the bond dimension $D$. If the Schmidt error remains roughly constant as we increase it polynomially , $\gep^{(N/2)}_{D=N^{\gb}}\approx \gep_{0}$, we can rule out weak logarithmic behavior by obtaining $S^{(n)}\ge \gb \gep_{0}\log N$. If the behavior at the critical point is known, and in particular if it is logarithmic with some known coefficient, one could show that the entropy is higher than that for a particular $p$. Alternately, the this information can be used as qualitative evidence for, especially if we find a sharp transition at the same $p$ we observe a break in the upper bound described above.

Notably, this lower bound estimate makes no use of $\F$; it relies entirely on feedback from the numerical estimate of the MPS. This obviously reduces the quantum overhead required, but it also reduces the classical computation somewhat, as a smaller light-cone style circuit can be calculated to extract only ${\epsilon^{(N/2)}_{D}}$. Still, this bound is logarithmic in $D$ and so the resource requirements scale unfavorably. However, as mentioned above, the upper bound, which has polynomial requirements, appears to be tight and may prove sufficient to extract the critical point $\pc$. Alternately, one may combine the upper bound from $\F$ with lower bounds estimated by some other means \cite{Garratt2023,McGinley2023}.

\pardash{Computational overhead and realization} We consider the computational resources required by our proposed protocol, both classical and quantum. The quantum resources required are outlined in \cref{fig:sketch}. These are polynomial in $N$ by construction: the application of the monitored random circuit requires a depth of $L\propto N$ layers, with each layer having $N/2$ concurrent two-qubit gates and $pN$ concurrent measurements. The inverted circuit can be straightforwardly decomposed into $N$ unitary $\log_{2}D$-qubit operations \cite{supplemental, Schon2005, Ran2020}, which can be further decomposed into $O(D^{2})$ two-qubit gates each \cite{Shende2006, Krol2022} or $\tilde O(\sqrt{D})$ gates and $O\p{D}$ ancillas \cite{Rosenthal2023}.

The classical computation to calculate the MPS at each gate consists of matrix multiplication and application of the two-qubit gate followed by a singular value decomposition, with complexity $O(D^{3})$, repeated $N/2$ times at each layer. This process is done sequentially in parallel with the quantum computation as output arrives, as each update of the MPS requires only the measurement results from a single layer. Alternately, the finished state would be held in quantum memory as the computation takes place.

\pardash{Conclusion and outlook} We've presented here a hybrid quantum algorithm for probing the entropy transition in monitored random circuits by generating a unitary mirror, showing a path to obtain a rigorous upper bound on the amount of entanglement entropy in a qubit chain, and so on the critical point $\pc$, with only polynomial quantum and classical resources.

Throughout the paper we have focused on the common case of a one-dimensional chain. In fact, the derivation described in \crefrange{eq:defen}{eq:ubSN} generalizes to higher dimensional systems. The number of terms in the Schmidt decomposition of the tensor network simply goes from $D\to D^{\abs{A}}$, where $\abs{A}$ is the number of qubits in the boundary of the subset, and the same bounds follow. However, it may not be straightforward to generate an appropriate high-fidelity MPS in polynomial time and convert it into a unitary mirror circuit with a polynomial number of gates. These may still be possible with specific forms of the tensor network \cite{Zaletel2020}.

While we focus here on pinpointing the volume law transition, the mirror fidelity may in itself have interesting properties \cite{Ware2023}. We have hypothesized about that the upper bound $\pcest$ is tight. If it is not, then perhaps there is a different critical point $\pc^{(\F)}$; a divergence of these could hint at some measure of entanglement in a state beyond the entanglement entropy which could be probed by the unitary mirror.

Alternatively, one can imagine an version of the scheme where the mirror circuit is applied directly to a set of auxiliary qubits, generating the state $\ket{\psi}\otimes\ket{\psi_{D}}$. This could be used as a resource in any number of generalized schemes to probe the monitored circuit ensemble \cite{McGinley2023}.

In addition, while our algorithm focuses on producing $\F$, the mirrors calculated can be stored beyond the initial run. Where we find $\F\sim 1$, they are good approximations of the real state, and so analyzing them can give us any insights we desire about the actual state produced by the monitored circuits. This may be true up to the critical point itself.

Finally, we note that the resources needed to produce a quantum-advantage result here are not large. The quantum resources required for the unitary mirror, in particular, may be as little as $D$ ancilla qubits and a depth of $O(\sqrt{D})$ gates \cite{supplemental, Rosenthal2023}. Thus, they are likely to be smaller than those required for the monitored circuit itself. The other roadblock to implementing such algorithms is the need to implement quantum-classical communication mid-circuit, and go beyond simple \emph{if} statements in the quantum code. Such capabilities would open the door to many more hybrid algorithms.

\bibliography{}
\widetext
\clearpage

\begin{center}
\textbf{\large Supplemental Material: Detecting Measurement-Induced Entanglement Transitions With Unitary Mirror Circuits}
\end{center}

\setcounter{equation}{0}
\setcounter{figure}{0}
\setcounter{table}{0}
\renewcommand{\theequation}{S\arabic{equation}}
\renewcommand{\thefigure}{S\arabic{figure}}
\renewcommand{\bibnumfmt}[1]{[S#1]}
\renewcommand{\citenumfont}[1]{S#1}

\section{Maximum fidelity of an MPS}

Let $\ket{\psi}$ be some wavefunction on $N$ qubits, given by a Schmidt decomposition,
\begin{equation}
\ket{\psi} = \sum_{i=1}^{2^{\bar n}}\sqrt{\mu_{i}}\ket{A_{i}}\ket{B_{i}}
\end{equation}
so that $\ket{A_{i}},\ket{B_{i}}$ described the state of qubits $1\dotsc n$, $n+1\dotsc N$, respectively, and have
\begin{equation*}\begin{gathered}
\mu_{i}\ge \mu_{i+1}, \qquad \sum_{i=1}^{2^{\bar n}}\mu_{i} = 1,
\\ \braket{A_{i}}{A_{j}} = \gd_{ij}, \qquad \braket{B_{i}}{B_{j}} = \gd_{ij}.
\end{gathered}\end{equation*}

The Schmidt error is given by
\begin{equation}
\en = \smashoperator{\sum_{i=D+1}^{2^{\bar n}}}\mu_{i}.
\end{equation}

Let $\ket{\psi_{D}}$ be an approximate wavefunction, whose Schmidt decomposition has only $D$ terms,
\begin{equation}
\ket{\psi_{D}} = \sum_{i=1}^{D}\sqrt{\gl_{i}}\ket{\ga_{i}}\ket{\gb_{i}}
\end{equation}
where
\begin{equation*}\begin{gathered}
\gl_{i}\ge \gl_{i+1}, \qquad \sum_{i=1}^{2^{\bar n}}\gl_{i} = 1,
\\ \ket{\ga_{i}} = \sum_{j}U_{ij}\ket{A_{j}}, \qquad \ket{\gb_{i}} = \sum_{j}V_{ij}\ket{B_{j}},
\end{gathered}\end{equation*}
and $U,V$ are two $2^{\bar n}\times2^{\bar n}$ unitary matrices.

The overlap between the two is given by
\begin{equation*}\begin{split}
\braket{\psi}{\psi_{D}} & = \sum_{i,j}\sqrt{\mu_{i}}\sqrt{\gl_{j}}\braket{A_{i}}{\ga_{j}}\braket{B_{i}}{\gb_{j}}
	\\ & = \sum_{i,j}\sqrt{\mu_{i}}\sqrt{\gl_{j}}U_{ij}V_{ij}
	\\ & = \Tr\br{\sqrt{\mu}\cdot U\cdot \sqrt{\gl}\cdot V^{T}}
\end{split}\end{equation*}
where $\sqrt{\mu},\sqrt{\gl}$ are diagonal $2^{\bar n}\times2^{\bar n}$. By the Von Neumann trace inequality, we have
\begin{equation}
\abs{\braket{\psi}{\psi_{D}}} \le \sum_{i=1}^{D}\sqrt{\gl_{i}}\sqrt{\mu_{i}};
\end{equation}
the right hand side is maximized by $\gl_{i} = \mu_{i}/\sum_{i=1}^{D}\mu_{i}$, leading to $\abs{\braket{\psi}{\psi_{D}}}^{2} \le \sum_{i=1}^{D}\mu_{i} = 1- \en$.

\section{Relations between $\en$ and the entropy}

We show here an upper and lower bound on the Von Neumann entanglement entropy of a subset of qubits based on the Schmidt error $\en$.

The R\'enyi entanglement entropy of order $\ga$ is given by
\begin{equation}
S^{(n)}_{\ga} = \frac{1}{1-\ga}\log\br{\Tr \rho_{A}^{\ga}} = \frac{1}{1-\ga}\log\br{\sum_{i=1}^{2^{\bar n}}\mu_{i}^{\ga}},
\end{equation}
while the Von Neumann entanglement entropy is given by
\begin{equation}
S^{(n)} = \lim_{\ga\to 1}S^{(n)}_{\ga} = -\sum_{i=1}^{2^{\bar n}}\mu_{i}\log \mu_{i}.
\end{equation}

\subsection{Extremizing the logand}

Define, for any $0\le p_{\min}\le \Sigma/m \le p_{\max} \le \Sigma - (m-1)p_{\min}$,
\begin{equation}\begin{split}
F^{\pm}_{\ga}\p{m,\Sigma,p_{\min},p_{\max}} & = \pm\sign\br{1-\ga} 
	\max_{\substack{\{\mu_{1},\dotsc,\mu_{m}\}\\ \text{so that}\\ \sum_{i=1}^{m} \mu_{i} = \Sigma \\ \forall \mu_{i}, \; p_{\min} \le \mu_{i} \le p_{\max}}}\br{\pm \sign\br{1-\ga}\sum_{i=1}^{m} \mu_{i}^{\ga}},
\end{split}\end{equation}
So that $F^{+}$ maximizes the R\'enyi entropy and $F^{-}$ minimizes it.

We prove by induction:
\begin{subequations}\begin{gather}
\label{eq:Fmax} F^{+}_{\ga}\p{m,\Sigma,p_{\min},p_{\max}} = m\p{\Sigma/m}^{\ga},
\\ \label{eq:Fmin0} F^{-}_{\ga}\p{m,\Sigma,0,p_{\max}} = \lfloor\frac{\Sigma}{p_{\max}}\rfloor(p_{\max})^{\ga} + \p{\Sigma - \lfloor\frac{\Sigma}{p_{\max}}\rfloor p_{\max}}^{\ga},
\\  \label{eq:FminSigma} F^{-}_{\ga}\p{m,\Sigma,p_{\min},\Sigma} = \p{\Sigma - \p{m-1}p_{\min}}^{\ga} + \p{m-1}(p_{\min})^{\ga}.
\end{gather}\end{subequations}

\subsubsection{Proof \ref{eq:Fmax}}

At $m=1$,
\begin{equation}\begin{gathered}
F^{+}_{\ga}\p{1,\Sigma,p_{\min},p_{\max}} = \Sigma^{\ga}.
\end{gathered}\end{equation}

Assume $F^{+}_{\ga}\p{m,\Sigma,p_{\min},p_{\max}} = m\p{\Sigma/m}^{\ga}$. Then
\begin{equation*}\begin{split}
F^{+}_{\ga}\p{m+1,\Sigma,p_{\min},p_{\max}} & = \sign\br{1-\ga} 
	\max_{\substack{\{\mu_{1},\dotsc,\mu_{m+1}\}\\ \sum_{i=1}^{m+1} \mu_{i} = \Sigma \\ p_{\min} \le \mu_{i} \le p_{\max}}}\br{\sign\br{1-\ga}\sum_{i=1}^{m+1} \mu_{i}^{\ga}}
	\\ & = \sign\br{1-\ga} \smashoperator{\max_{\substack{{\bar \mu}\\\frac{\Sigma}{m+1} \le {\bar \mu}\le p_{\max}}}}\Big[{\sign\br{1-\ga}{\bar \mu}^{\ga} + 
	\max_{\substack{\{\mu_{2},\dotsc,\mu_{m+1}\}\\ \sum_{i=2}^{m+1} \mu_{i} = \Sigma - {\bar \mu} \\ p_{\min} \le \mu_{i} \le {\bar \mu}}}\br{\sign\br{1-\ga}\sum_{i=2}^{m+1} \mu_{i}^{\ga}}}\Big]
	\\ & = \sign\br{1-\ga} \smashoperator{\max_{\substack{{\bar \mu}\\\frac{\Sigma}{m+1} \le {\bar \mu}\le p_{\max}}}}\Big[{\sign\br{1-\ga}\p{{\bar \mu}^{\ga} + F^{+}_{\ga}\p{m,\Sigma-{\bar \mu},p_{\min},{\bar \mu}}}}\Big]
	\\ & = \sign\br{1-\ga} \smashoperator{\max_{\substack{{\bar \mu}\\\frac{\Sigma}{m+1} \le {\bar \mu}\le p_{\max}}}}\Big[{\sign\br{1-\ga}\p{{\bar \mu}^{\ga} + m\br{\frac{\Sigma-{\bar \mu}}{m}}^{\ga}}}\Big].
\end{split}\end{equation*}
Here, we separated the term in the brackets into the largest coefficient $\bar \mu$ and the rest, and then identified the remaining sum as the maximized function for $m$, and inserted its value as per the induction assumption.

Maximizing this, we see
\begin{equation*}\begin{gathered}
\dee{}{{\bar \mu}}\br{\sign\br{1-\ga}\p{{\bar \mu}^{\ga} + m\br{\frac{\Sigma-{\bar \mu}}{m}}^{\ga}}} = \sign\br{1-\ga}\ga\p{ {\bar \mu}^{\ga-1} -  \br{\p{\Sigma-{\bar \mu}}/m}^{\ga-1}} < 0\quad\text{for }{\bar \mu}>\Sigma/(m+1),
\\ \Rightarrow F^{+}_{\ga}\p{m+1,\Sigma,p_{\min},p_{\max}} = \p{m+1}\p{\frac{\Sigma}{m+1}}^{\ga}.
\end{gathered}\end{equation*}

\subsubsection{Proof \ref{eq:Fmin0}}

At $m=1$,
\begin{equation}\begin{gathered}
F^{-}_{\ga}\p{1,\Sigma,0,p_{\max}} = \Sigma^{\ga}.
\end{gathered}\end{equation}

Assume $F^{-}_{\ga}\p{m,\Sigma,0,p_{\max}} = \lfloor\frac{\Sigma}{p_{\max}}\rfloor(p_{\max})^{\ga} + \p{\Sigma - \lfloor\frac{\Sigma}{p_{\max}}\rfloor p}^{\ga}$. Then, repeating the same procedure,
\begin{equation*}\begin{split}
F^{-}_{\ga}\p{m+1,\Sigma,0,p_{\max}} & = \sign\br{1-\ga} 
	\min_{\substack{\{\mu_{1},\dotsc,\mu_{m+1}\}\\ \sum_{i=1}^{m+1} \mu_{i} = \Sigma \\ 0 \le \mu_{i} \le p_{\max}}}\br{\sign\br{1-\ga}\sum_{i=1}^{m+1} \mu_{i}^{\ga}}
	\\ & = \sign\br{1-\ga} \smashoperator{\min_{\substack{{\bar \mu}\\\frac{\Sigma}{m+1} \le {\bar \mu}\le p_{\max}}}}\Big[{\sign\br{1-\ga}{\bar \mu}^{\ga} + 
		\min_{\substack{\{\mu_{2},\dotsc,\mu_{m+1}\}\\ \sum_{i=2}^{m+1} \mu_{i} = \Sigma - {\bar \mu} \\ 0 \le \mu_{i} \le {\bar \mu}}}\br{\sign\br{1-\ga}\sum_{i=2}^{m+1} \mu_{i}^{\ga}}}\Big]
	\\ & = \sign\br{1-\ga} \smashoperator{\min_{\substack{{\bar \mu}\\\frac{\Sigma}{m+1} \le {\bar \mu}\le p_{\max}}}}\Big[{\sign\br{1-\ga}\p{{\bar \mu}^{\ga} + F^{-}_{\ga}\p{m,\Sigma-{\bar \mu},0,{\bar \mu}}}}\Big]
	\\ & = \sign\br{1-\ga} \smashoperator{\min_{\substack{{\bar \mu}\\\frac{\Sigma}{m+1} \le {\bar \mu}\le p_{\max}}}}\Big[{\sign\br{1-\ga}\p{{\bar \mu}^{\ga} + 
		\lfloor\frac{\Sigma - {\bar \mu}}{{\bar \mu}}\rfloor {\bar \mu}^{\ga} + \br{\Sigma - {\bar \mu} - \lfloor\frac{\Sigma - {\bar \mu}}{{\bar \mu}}\rfloor {\bar \mu}}^{\ga}}}\Big]
	\\ & = \sign\br{1-\ga} \smashoperator{\min_{\substack{{\bar \mu}\\\frac{\Sigma}{m+1} \le {\bar \mu}\le p_{\max}}}}
		\Big[{\sign\br{1-\ga}\p{\lfloor\frac{\Sigma }{{\bar \mu}}\rfloor {\bar \mu}^{\ga} + \br{\Sigma - \lfloor\frac{\Sigma}{{\bar \mu}}\rfloor {\bar \mu}}^{\ga}}}\Big].
\end{split}\end{equation*}

Minimizing this, for any integer $\ell$,
\begin{equation*}\begin{split}
\dee{}{{\bar \mu}}& \br{\sign\br{1-\ga}\p{\lfloor\frac{\Sigma }{{\bar \mu}}\rfloor {\bar \mu}^{\ga} + \br{\Sigma - \lfloor\frac{\Sigma}{{\bar \mu}}\rfloor {\bar \mu}}^{\ga}}}_{\Sigma/{\bar \mu}\notin \mathbb N}  = 
	\\ & \quad  \sign\br{1-\ga}\br{\ga \lfloor\frac{\Sigma }{{\bar \mu}}\rfloor {\bar \mu}^{\ga-1}\p{1 - \br{\frac{\Sigma}{{\bar \mu}} - \lfloor\frac{\Sigma}{{\bar \mu}}\rfloor }^{\ga-1}}} < 0
\\ & \br{\sign\br{1-\ga}\p{\lfloor\frac{\Sigma}{{\bar \mu}}\rfloor {\bar \mu}^{\ga} + \br{\Sigma - \lfloor\frac{\Sigma}{{\bar \mu}}\rfloor {\bar \mu}}^{\ga}}}_{{\bar \mu}=\Sigma/\ell + \gve} 
	- \br{\sign\br{1-\ga}\p{\lfloor\frac{\Sigma}{{\bar \mu}}\rfloor {\bar \mu}^{\ga} + \br{\Sigma - \lfloor\frac{\Sigma}{{\bar \mu}}\rfloor {\bar \mu}}^{\ga}}}_{{\bar \mu}=\Sigma/\ell} = 
\\ & \quad \sign\br{1-\ga}\br{\p{\ell-1}\p{\Sigma/\ell + \gve}^{\ga} + \br{\Sigma - \p{\ell-1}\p{\frac{\Sigma}{\ell} + \gve}}^{\ga} - \ell\p{\Sigma/\ell}^{\ga}}
\\ & \quad = -\ga\abs{1-\ga}\p{\Sigma/\ell}^{\ga-2}\ell \p{\ell-1}\gve^{2}/2 + O\p{\gve}^{3} < 0 
\end{split}\end{equation*}
we find
\begin{equation*}\begin{split}
F^{-}_{\ga}\p{m+1,\Sigma,0,p_{\max}} & = \lfloor\frac{\Sigma }{p_{\max}}\rfloor (p_{\max})^{\ga} + \p{\Sigma - \lfloor\frac{\Sigma}{p_{\max}}\rfloor p_{\max}}^{\ga}.
\end{split}\end{equation*}

\subsubsection{Proof \ref{eq:FminSigma}}

At $m=1$,
\begin{equation}\begin{gathered}
F^{-}_{\ga}\p{1,\Sigma,p_{\min},\Sigma} = \Sigma^{\ga}.
\end{gathered}\end{equation}

Assume $F^{-}_{\ga}\p{m,\Sigma,p_{\min},\Sigma} = \p{\Sigma - \p{m-1}p_{\min}}^{\ga} + \p{m-1}(p_{\min})^{\ga}$. Then
\begin{equation*}\begin{split}
F^{-}_{\ga}\p{m+1,\Sigma,p_{\min},\Sigma} & = \sign\br{1-\ga} 
	\min_{\substack{\{\mu_{1},\dotsc,\mu_{m+1}\}\\ \sum_{i=1}^{m+1} \mu_{i} = \Sigma \\ p_{\min} \le \mu_{i} \le \Sigma}}\br{\sign\br{1-\ga}\sum_{i=1}^{m+1} \mu_{i}^{\ga}}
	\\ & = \sign\br{1-\ga} \smashoperator{\min_{\substack{{\bar \mu} \\ p_{\min}\le {\bar \mu} \le \frac{\Sigma}{m+1}}}}\Big[{\sign\br{1-\ga}{\bar \mu}^{\ga} + 
		\min_{\substack{\{\mu_{1},\dotsc,\mu_{m}\}\\ \sum_{i=1}^{m} \mu_{i} = \Sigma - {\bar \mu} \\ {\bar \mu} \le \mu_{i} \le \Sigma-{\bar \mu}}}\br{\sign\br{1-\ga}\sum_{i=1}^{m} \mu_{i}^{\ga}}}\Big]
	\\ & = \sign\br{1-\ga} \smashoperator{\min_{\substack{{\bar \mu} \\ p_{\min}\le {\bar \mu} \le \frac{\Sigma}{m+1}}}}\Big[{\sign\br{1-\ga}\p{{\bar \mu}^{\ga} + 
		F^{-}_{\ga}\p{m,\Sigma-{\bar \mu},{\bar \mu},\Sigma-{\bar \mu}}}}\Big]
	\\ & = \sign\br{1-\ga} \smashoperator{\min_{\substack{{\bar \mu} \\ p_{\min}\le {\bar \mu} \le \frac{\Sigma}{m+1}}}}\Big[{\sign\br{1-\ga}\p{{\bar \mu}^{\ga} + 
		\p{\Sigma - {\bar \mu} - \p{m-1}{\bar \mu}}^{\ga} + \p{m-1}{\bar \mu}^{\ga}}}\Big]
	\\ & = \sign\br{1-\ga} \smashoperator{\min_{\substack{{\bar \mu} \\ p_{\min}\le {\bar \mu} \le \frac{\Sigma}{m+1}}}}\Big[{\sign\br{1-\ga}\p{\br{\Sigma - m{\bar \mu}}^{\ga} + m{\bar \mu}^{\ga}}}\Big].
\end{split}\end{equation*}
This time we have separated the smallest term $\bar \mu$.

Minimizing,
\begin{equation*}\begin{gathered}
\dee{}{\bar \mu} \br{\sign\br{1-\ga}\p{\br{\Sigma - m\bar \mu}^{\ga} + m{\bar \mu}^{\ga}}}  = m\ga \sign\br{1-\ga}\p{{\bar \mu}^{\ga-1} - \br{\Sigma - m\bar \mu}^{\ga-1}} > 0
\\ \Rightarrow F^{-}_{\ga}\p{m+1,\Sigma,p_{\min},\Sigma} = \p{\Sigma - mp_{\min}}^{\ga} + m(p_{\min})^{\ga}.
\end{gathered}\end{equation*}

\subsection{Upper bound}

We have now
\begin{equation}
S^{(n)}_{\ga}\p{\rho_{n}} = \frac{1}{1-\ga}\log\br{\sum_{i=1}^{2^{\bar n}}\mu_{i}^{\ga}},
\end{equation}
with
\begin{equation*}\begin{gathered}
1- \en  = {\sum_{i=1}^{D}\mu_{i}^{\ga}}, \qquad \en = {\sum_{i=D+1}^{2^{\bar n}}\mu_{i}^{\ga}}, 
\\ \mu_{i} > \mu_{i+1}
\end{gathered}\end{equation*}

Thus, we have for the upper bound,
\begin{equation}\begin{split}
S^{(n)}_{\ga}  & \le  \max_{\mu_{D}}\frac{1}{1-\ga}\log\br{F^{+}\p{D,1-\en,\mu_{D},1-\en-(D-1)\mu_{D}} + F^{+}\p{2^{\bar n}-D, \en, 0, \mu_{D}}}
	\\ & =  \max_{\mu_{D}}\frac{1}{1-\ga}\log\br{D^{1-\ga}\p{1-\en}^{\ga} + \p{2^{\bar n} - D}^{1-\ga}\p{\en}^{\ga}}.
	\\ & = \frac{1}{1-\ga}\log\br{D^{1-\ga}\p{1-\en}^{\ga} + \p{2^{\bar n} - D}^{1-\ga}\p{\en}^{\ga}}.
\end{split}\end{equation}
and taking the limit $\ga\to1$, we find
\begin{equation}
S^{(n)} \le -\en\log\br{\en} - \p{1-\en}\log\br{1-\en} + \en\log\br{2^{\bar n} - D} + \p{1-\en}\log D.
\end{equation}

We can further bound this, as
\begin{equation}\begin{split}
S^{(n)} & \le \log 2 + \en\log\br{2^{\bar n}} + \p{1-\en}\log D \\ & = \log 2 + \en\log\br{2^{\bar n}/D} + \log D
	\\ & \le \log 2 + \p{1-F}\log\br{2^{\bar n}/D} + \log D \\ & = \log 2 + \bar n\p{1-F}\log2 + F\log D.
\end{split}\end{equation}
Using first $0\le\en\le1$ and then $D\le 2^{\bar n}$ and $\en\le 1-F$. With $\bar n \le N/2$ we find the bound used in the main text.

\subsection{Lower bound}

Similarly, for the lower bound,
\begin{equation}\begin{split}
S^{(n)}_{\ga} & \ge \min_{\mu_{D}}\frac{1}{1-\ga}\log\br{F^{-}\p{D,1-\en,\mu_{D},1-\en-(D-1)\mu_{D}} + F^{-}\p{2^{\bar n}-D, \en, 0, \mu_{D}}}
\\ & = \min_{\mu_{D}}\frac{1}{1-\ga}\log\br{\p{1 - \en - \p{D-1}\mu_{D}}^{\ga} + \p{D-1}(\mu_{D})^{\ga} + 
	\lfloor\frac{\en }{\mu_{D}}\rfloor (\mu_{D})^{\ga} + \p{\en - \lfloor\frac{\Sigma}{\mu_{D}}\rfloor \mu_{D}}^{\ga}}
\\ & \approx \min_{\mu_{D}}\frac{1}{1-\ga}\log\br{\p{1 - \en - \p{D-1}\mu_{D}}^{\ga} + \p{\en + \p{D-1}\mu_{D}}(\mu_{D})^{\ga-1}}.
\end{split}\end{equation}
By convexity,
\begin{equation*}\begin{split}
& \frac{1}{1-\ga}\log\br{\p{1 - \en - \p{D-1}\mu_{D}}^{\ga} + \p{\en + \p{D-1}\mu_{D}}(\mu_{D})^{\ga-1}} \ge 
	\\ & \frac{1}{1-\ga}\log\br{\p{1 - \en}^{\ga} - \br{\p{1 - \en}^{\ga} - \p{1 - \en - \p{D-1}\mu_{D}}^{\ga}}\p{\frac{\mu_{D}}{\mu_{D}^{\max}}}^{\ga} + \p{\en + \p{D-1}\mu_{D}}(\mu_{D})^{\ga-1}},
\end{split}\end{equation*}
and inserting $\mu_{D}^{\max} = \p{1-\en}/D$, one finds
\begin{equation}\begin{split}
S^{(n)}_{\ga} & \ge \min_{\mu_{D}}\frac{1}{1-\ga}\log\br{\p{1 - \en}^{\ga} + \br{\frac{D - D^{\ga}}{1-\ga}}^{1-\ga}\p{\frac{\en}{\ga}}^{\ga}}.
\end{split}\end{equation}

Once again, taking the limit $\ga \to 1$, we find
\begin{equation}
S^{(n)} \ge -\en\log\br{\en} - \p{1-\en}\log\br{1-\en} + \en\p{1 + \log\br{ D\log D}}.
\end{equation}

\section{Calculation of the MPS}

Here we describe the time-evolving block decimation (TEBD) algorithm used to generate the MPS approximation of the quantum state. 

Roughly speaking, we absorb all measurement projection operators into the preceding gates to generate a set of (non-unitary) two-qubit operators. To apply an operator, we renormalize the MPS into a canonical form with center position at one of the two qubits, calculate the tensor multiplication, and then perform a singular value decomposition (SVD) and concatenate down to $D$ singular value. We apply these operators layer by layer in a snaking pattern to reduce the overhead associated with the canonical renormalization.

Formally, we are given as an input:
\begin{itemize}
\item A set of unitary gates $\setU = \{\hat U_{\ell,2q} | 1\le\ell\le L, 1\le q\le \lfloor N/2\rfloor\}$, so that at each odd (even) layer $\ell$, $\hat U_{\ell,q}$ is applied between each even qubit $2q$ and its preceding (following) neighbor, $2q+\p{-1}^{\ell}$,
\item A set of measurement points $\setM = \{M_{\ell}|1\le\ell\le L\}$, where $M_{\ell}\subset\{1,\dotsc,N\}$, specifying for each layer $\ell$ which qubits were measured, and
\item A measurement record, $\setm \in \left\{m_{\ell,q} | 1\le\ell\le L, q\in M_{\ell} \right\}$, where $m_{\ell,q}\in \left\{0,1\right\}$ is the result of the measurement of qubit $q$ at layer $\ell$.
\end{itemize}

The MPS is given by
\begin{equation}
\ket{\psi_{D}} = \smashoperator{\sum_{\substack{i_{0},\dotsc,i_{N},\\x_{1},\dotsc,x_{N}}}}
	A^{(1)}_{i_{0}x_{1}i_{1}}\dotsi A^{(N)}_{i_{N-1}x_{1}i_{N}}\ket{x_{1}\dotsc x_{N}}
	\label{eq:MPSdef}
\end{equation}
\begin{itemize}
\item A bond dimension $D$;
\item $N$ tensors, $A^{(n)}_{i_{n-1}x_{q}i_{n}}$, with $n=1,\dotsc,N$, $x_{{n}}=0,1$, $i_{n} = 1,\dotsc,\min\br{2^{n},2^{N-n},D}$;
\item A center position $1 \le n_{c} \le N$, so that for any $n>n_{c}$ $A^{(n)}$ is right-orthogonal, $\sum_{x,j}A^{(n)}_{ixj}A^{(n)*}_{i\pr xj} = \gd_{ii\pr}$, for any $n<n_{c}$ it is left-orthogonal, $\sum_{i,x}A^{(n)}_{ixj}A^{(n)*}_{ixj\pr} = \gd_{jj\pr}$, and $\sum_{i,x,j}A^{(n_{c})}_{ixj}A^{(n_{c})*}_{ixj} = 1$.
\end{itemize}

Our algorithm proceeds as follows:
\begin{enumerate}
	\item Initialize all tensors to $A^{(n)}_{ixj} \to \gd_{i,1}\gd_{j,1}\gd_{x,0}$, so that the inital state is $\ket{\psi_{D}} = \ket{0\dotsi 0}$
	\item For each layer $\ell = 1,\dotsc,\L$,
	\begin{enumerate}[label*=\arabic*.]
		\item If $\ell$ is even, and $N$ is even, and $N\in M_{\ell}$, apply a projection $\ket{m_{\ell,N}}\bra{m_{\ell_{N}}}$ to $A^{(N)}$:
		\begin{enumerate}[label*=\arabic*.]
			\item Renormalize the MPS to set the center position $n_{c} = N$
			\item Set $A^{(N)}_{ixj} \to \gd_{x,m_{\ell,N}}A^{(N)}_{ixj}/\sqrt{\sum_{i\pr j\pr}\abs{A^{(N)}_{i\pr m_{\ell,N}j\pr}}^{2}}$
		\end{enumerate}
		\item For each:
		\begin{itemize}
			\item $q_{1} = 1,3,5,\dotsc N-1$ if $\ell$ is odd,
			\item $q_{1} = N-2, N-4, \dotsc, 2$ if $\ell$ is even and $N$ is even,
			\item $q_{1} = N-1, N-3, \dotsc, 2$ if $\ell$ is even and $N$ is odd:
		\end{itemize}
		\begin{enumerate}[label*=\arabic*.]
			\item Renormalize the MPS to set the center position $n_{c} = q_{1}$
			\item Let $q_{2} = q_{1} + 1$
			\item If $q_{1}\in M_{\ell}$, let $\hat P_{1} = \ket{m_{\ell,q_{1}}}\bra{m_{\ell,q_{1}}}$; otherwise let $\hat P_{1} = 1$
			\item If $q_{2}\in M_{\ell}$, let $\hat P_{2} = \ket{m_{\ell,q_{2}}}\bra{m_{\ell,q_{2}}}$; otherwise let $\hat P_{2} = 1$
			\item Apply $\p{\hat P_{1}\otimes \hat P_{2}} \hat U_{\ell,2i}$ to qubits $q_{1},q_{2}$:
			\begin{enumerate}[label*=\arabic*.]
				\item Let $B_{ixyj} = \sum_{x\pr,y\pr}\bra{xy}\p{\hat P_{1}\otimes \hat P_{2}} \hat U_{\ell,2i}\ket{x\pr y\pr}A^{(q_{1})}_{ix\pr l}A^{(q_{2})}_{ly\pr j}$
				\item Perform SVD, $B_{ixyj} = \sum_{l}V_{ixl}\gl_{l}W_{lyj}$, so that $V$ ($W$) is left (right) orthogonal
				\item Discard the smaller singular values, keeping at most $D$ terms, $l\le D$
				\item Set $A^{(q_{1})}_{ixj} \to V_{ixj}\gl_{j}/\sqrt{\sum_{i\pr x\pr j\pr}\abs{V_{i\pr x\pr j\pr }\gl_{j}\pr }^{2}}$ 
				\item Set $A^{(q_{2})}_{ixj} \to W_{ixj}$
			\end{enumerate}
		\end{enumerate}
		\item If $\ell$ is odd, and $N$ is odd, and $N\in M_{\ell}$, apply a projection $\ket{m_{\ell,N}}\bra{m_{\ell_{N}}}$ to $A^{(N)}$
		\item If $\ell$ is even, and $1\in M_{\ell}$, apply a projection $\ket{m_{\ell,1}}\bra{m_{\ell_{1}}}$ to $A^{(1)}$
	\end{enumerate}
\end{enumerate}

\section{Circuit generating the Unitary Mirror}

We give the circuit generating the unitary mirror. Let the MPS be defined as in \cref{eq:MPSdef}, and let $\nD = \lceil\log_{2}D\rceil$. We define the following unitaries:
\begin{itemize}
\item For $n_{c}$, we define $U^{(n_{c})}$ as any ($2\nD+1$)-qubit circuit that has
\begin{equation*}
U^{(n_{c})}\ket{0^{\otimes\nD}}\otimes\ket{0\otimes\ket{0}^{\otimes\nD}} = \sum_{i,x,j}A^{(n_{c})}_{ixj}\ket{i}\otimes \ket{x}\otimes\ket{j}.
\end{equation*}
As $\sum_{i,x,j}A^{(n_{c})}_{ixj}A^{(n_{c})*}_{ixj} = 1$ this is a valid unitary.

\item For $N - \nD >n > n_{c}$, we define $U^{(n)}$ as any ($\nD+1$)-qubit circuit that has
\begin{equation*}
U^{(n)}\ket{i}\otimes\ket{0} = \sum_{x,j}A^{(n)}_{ixj} \ket{x}\otimes\ket{j},
\end{equation*}
As $\sum_{x,l}A^{(n)}_{ixj}A^{(n)*}_{i\pr xj} = \gd_{ii\pr}$ this is a valid unitary.

\item For $1+\nD<n < n_{c}$, we define $U^{(n)}$ as any ($\nD+1$)-qubit circuit that has
\begin{equation*}
U^{(n_{c})}\ket{0}\otimes\ket{j} = \sum_{i,x}A^{(n)}_{ixj}\ket{i}\otimes \ket{x}.
\end{equation*}
As $\sum_{i,x}A^{(n_{c})}_{ixj}A^{(n_{c})*}_{ixj\pr} = \gd_{j,j\pr}$ this is a valid unitary.

\item Finally, for the edges, we define
\begin{equation*}\begin{gathered}
U^{(N-\nD)}\ket{i}\otimes\ket{0} = \smashoperator{\sum_{\substack{x_{N-\nD},\dotsc,x_{N},\\ j_{N-\nD},\dotsc,j_{N-1}}}}
	A^{(N-\nD)}_{ixj_{N-\nD}} \dotsc A^{(N)}_{j_{N-1}x_{N}} \ket{x_{N-\nD}\dotsc x_{N}},
\\ U^{(\nD+1)}\ket{0}\otimes\ket{j} = \smashoperator{\sum_{\substack{x_{1},\dotsc,x_{\nD+1},\\ i_{1},\dotsc,i_{\nD}}}}A^{(1)}_{xi_{1}} \dotsc A^{(\nD+1)}_{i_{\nD}x_{N}j} \ket{x_{1}\dotsc x_{\nD+1}}.
\end{gathered}\end{equation*}

\end{itemize}

\begin{figure}[htbp] 
   \centering
   \includegraphics[]{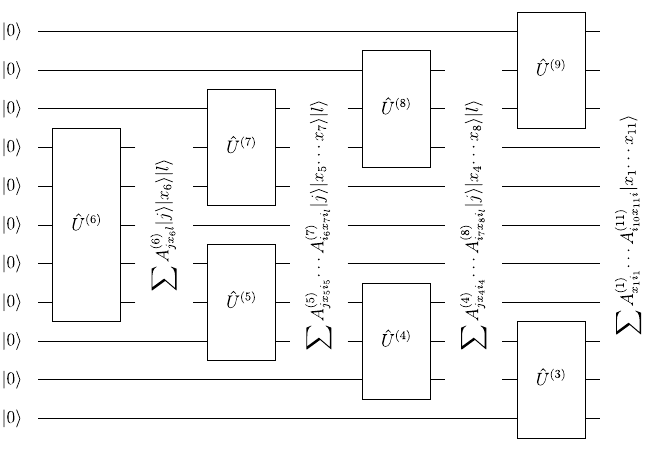} 
   \caption{An example unitary mirror circuit for $N=11$, $D=4$, $n_{c}=6$. The 5-qubit and 3-qubit gates shown here would be decomposed into a series of 2-qubit gates, depending the particular quantum machine used.}
   \label{fig:UMcirc}
\end{figure}

We then apply the gates sequentially to the all-zero state, starting with $U^{(n_{c})}$,
\begin{equation}
\ket{\psi}_{0} = \br{U^{(n_{c})}}_{n_{c}-\nD\dotsc n_{c}+\nD}\ket{0^{\otimes N}} = \smashoperator{\sum_{j,l,x_{n_{c}}}}A^{(n_{c})}_{jx_{n_{c}}l}
	\ket{0^{\otimes(\dotsi)}}\otimes \ket{j}\otimes \ket{x_{n_{c}}}\otimes\ket{l}\otimes \ket{0^{\otimes(\dotsi)}},
\end{equation}
and proceeding in both directions,
\begin{equation}\begin{split}
\ket{\psi}_{n} & = \br{U^{(n_{c})-n}}_{n_{c}-n-\nD \dotsc n_{c}-n}\br{U^{(n_{c})+n}}_{n_{c}+n \dotsc n_{c}+n+\nD}\ket{\psi}_{n-1} 
	\\ & = \smashoperator{\sum_{\substack{i_{n_{c}-n},\dotsc,i_{n_{c}+n-1},j,l\\x_{n_{c}-n},\dotsc,x_{n_{c}+n}}}}A^{(n_{c}-n)}_{jx_{n_{c}-n}i_{n_{c}-n}}\dotsi A^{(n_{c}+n)}_{i_{n_{c}+n-1}x_{n_{c}+n}l}
	\ket{0^{\otimes(\dotsi)}}\otimes \ket{j}\otimes \ket{x_{n_{c}-n}\dotsc x_{n_{c}+n}}\otimes\ket{l}\otimes \ket{0^{\otimes(\dotsi)}},
\end{split}\end{equation}
to arrive at the full state of \cref{eq:MPSdef} once all unitaries have been applied. The notation $\br{\hat U}_{i\dotsc j}$ implies applying $\hat U$ to qubits $i$ through $j$. The optimal circuit has $n_{c} = N/2$. An example of such a circuit is shown in \cref{fig:UMcirc}.

The inverted mirror takes the opposite shape, starting with gates on the outermost qubits and working inwards, followed by measurement of all qubits. The fidelity is obtained from the probability find all-zeros as the final state.

\end{document}